\documentclass[openacc]{rstransa}
\usepackage{siunitx}

\begin{document}

\title{FLASHForward: Plasma-wakefield accelerator science for high-average-power applications}

\author{R.~D'Arcy$^1$, A.~Aschikhin$^{1,2}$, S.~Bohlen$^{1,2}$, G.~Boyle$^1$, T.~Br{\"u}mmer$^1$, J.~Chappell$^{3}$, S.~Diederichs$^{2}$, B.~Foster$^{1,2,4}$, M.J.~Garland$^1$, L.~Goldberg$^{1,2}$, P.~Gonzalez$^{1,2}$, S.~Karstensen$^1$, A.~Knetsch$^1$, P.~Kuang$^1$, V.~Libov$^{1,2}$, K.~Ludwig$^1$, A.~Martinez de la Ossa$^{1,2}$, F.~Marutzky$^1$, M.~Meisel$^{1,2}$, T.J.~Mehrling$^{1,5}$, P.~Niknejadi$^1$, K.~Poder$^1$, P.~Pourmoussavi$^1$, M.~Quast$^{1,2}$, J.-H.~R{\"o}ckemann$^{1,2}$, L.~Schaper$^1$, B.~Schmidt$^1$, S.~Schr{\"o}der$^{1,2}$, J.-P.~Schwinkendorf$^{1,2}$, B.~Sheeran$^{1,2}$, G.~Tauscher$^{1,2}$, S.~Wesch$^1$, M.~Wing$^{1,3}$, P.~Winkler$^{1,2}$, M.~Zeng$^1$ and J.~Osterhoff$^1$
}

\address{$^1$Deutsches Elektronen-Synchrotron DESY, Notkestra{\ss}e 85, 22607 Hamburg, Germany\\
$^2$Universit{\"a}t Hamburg, Luruper Chaussee 149, 22761 Hamburg, Germany\\
$^3$University College London, Gower Street, London WC1E 6BT, UK\\
$^4$University of Oxford, Wellington Square, Oxford, OX1 2JD, UK\\
$^5$Lawrence Berkeley National Laboratory, University of California, Berkeley, California 94720, USA
}


\keywords{plasma wakefield acceleration, electrons, high average power}

\corres{Richard D'Arcy\\
\email{richard.darcy@desy.de}}

\newpage








\maketitle

\section*{Abstract}
The FLASHForward experimental facility is a high-performance test-bed for precision plasma-wakefield research, aiming to accelerate high-quality electron beams to GeV-levels in a few centimetres of ionised gas. The plasma is created by ionising gas in a gas cell either by a high-voltage discharge or a high-intensity laser pulse. The electrons to be accelerated will either be injected internally from the plasma background or externally from the FLASH superconducting RF front end. In both cases the wakefield will be driven by electron beams provided by the FLASH gun and linac modules operating with a 10~Hz macro-pulse structure, generating 1.25~GeV, 1~nC electron bunches at up to 3~MHz micro-pulse repetition rates. At full capacity, this FLASH bunch-train structure corresponds to 30~kW of average power, orders of magnitude higher than drivers available to other state-of-the-art LWFA and PWFA experiments. This high-power functionality means FLASHForward is the only plasma-wakefield facility in the world with the immediate capability to develop, explore, and benchmark high-average-power plasma-wakefield research essential for next-generation facilities. The operational parameters and technical highlights of the experiment are discussed, as well as the scientific goals and high-average-power outlook.

\section{Introduction}

In recent years, much focus and effort has been directed towards both cutting-edge particle-beam-driven (PWFA) \cite{Veksler,Chen} and laser-driven (LWFA) \cite{Tajima} plasma-wakefield-acceleration schemes. Huge strides have been made in both fields, with GV/m-level gradients demonstrated in both LWFA \cite{dreambeam,BELLA1} and PWFA \cite{FACET1,FACET2} experiments. In such schemes, either a particle beam or a high-intensity laser is injected into an ionised gas, i.e.~a plasma, displacing the plasma electrons due to space-charge fields and thus forming a bubble of effectively static ions in its wake. The electrostatic forces attract the displaced electrons back towards the positive ions left behind, with the electrons then oscillating around the central longitudinal axis of the impinging particle or ultra-intense laser beam. In the case where the density of the driving bunch is much greater than that of the plasma, also known as the blow-out regime due to the complete expulsion of electrons in the wake, the accelerating gradients experienced in PWFA schemes can reach $\mathcal{O}(100~{\rm GV/m})$ -- 1000~times larger than those of conventional RF accelerators \cite{Esarey2009}. These extreme gradients promise a dramatic reduction in accelerator footprint, drastically reducing the cost of future collider and photon-science facilities.

Due to the relative affordability of laser systems, LWFA has formed the backbone of plasma-wakefield experiments over the last few decades. Owing to the ubiquity of such laser facilities LWFA research has made great strides. Beam-driven experiments, in contrast, have been much less evident due to the scarcity of accessible machines with beam properties necessary for PWFA studies. Despite this PWFA is of continued interest due to its intrinsic advantages over LWFA:

\begin{itemize}
\item PWFA drivers propagate through plasma at close to the speed of light, $c$, whereas laser pulses travel at their group velocity -- lower than the speed of light. This increased velocity mitigates the tendency of the laser spot and the beam to drift gradually out of phase (dephasing).
\item The strong transverse focusing fields in PWFA schemes limit the expansion of the driving beam during propagation. This leads to much longer acceleration distances than those typical for unguided LWFA schemes, the utilisation of which is necessary for driver depletion and energy-transfer-efficiency studies.
\item Conventional accelerator sources are capable of producing high-average-power beams on the MW-level, essential for investigation of the applicability of plasma-wakefield schemes to future facilities. In contrast, state-of-the-art high-intensity laser systems deliver powers on the 100~W-level.
\end{itemize}

The FLASH free-electron laser (FEL) user facility \cite{FLASH}, capable of providing stable high-quality GeV-level electron bunches at MHz repetition rates with fs-level laser-to-electron-beam timing, fulfils all of these technical requirements. As such, the FLASHForward experiment (Future ORiented Wakefield Accelerator Research and Development at FLASH) \cite{FLASHForward} has been built to exploit these excellent beam properties, providing a location for cutting-edge high-precision PWFA studies. In addition, the FLASH superconducting radiofrequency (SRF) modules are capable of providing kW-level average-power beams to FLASHForward, making it the only facility in the world with the potential to explore this exciting regime of plasma-wakefield research.

In the following sections, the FLASHForward experimental facility will be presented in detail. Section~\ref{sec:FLASHForward} will provide an overview of the facility, its individual components, and outlook for future hardware upgrades. This will be followed in Sec.~\ref{sec:research} by a summary of the ongoing and planned scientific research at FLASHForward. In Sec.~\ref{sec:HAP}, conceptual possibilities of high-average-power PWFA experimentation at FLASHForward will be explained.


\section{The FLASHForward experimental facility}\label{sec:FLASHForward}

The FLASHForward experimental facility \cite{FLASHForward} is a high-performance test-bed for precision PWFA research (see Fig.~\ref{FIGURE:FLASH} for a schematic of the facility). Initial acceleration is carried out by the FLASH SRF stage using superconducting niobium accelerating cavities, providing high-quality, low-emittance, GeV-level electron bunches to the FLASHForward beamline. In addition to the supply of these high-quality bunches the FLASHForward facility also provides the opportunity to exploit the following technical highlights for innovative PWFA studies:

\begin{figure}[b]
\centering
\includegraphics[width=\columnwidth]{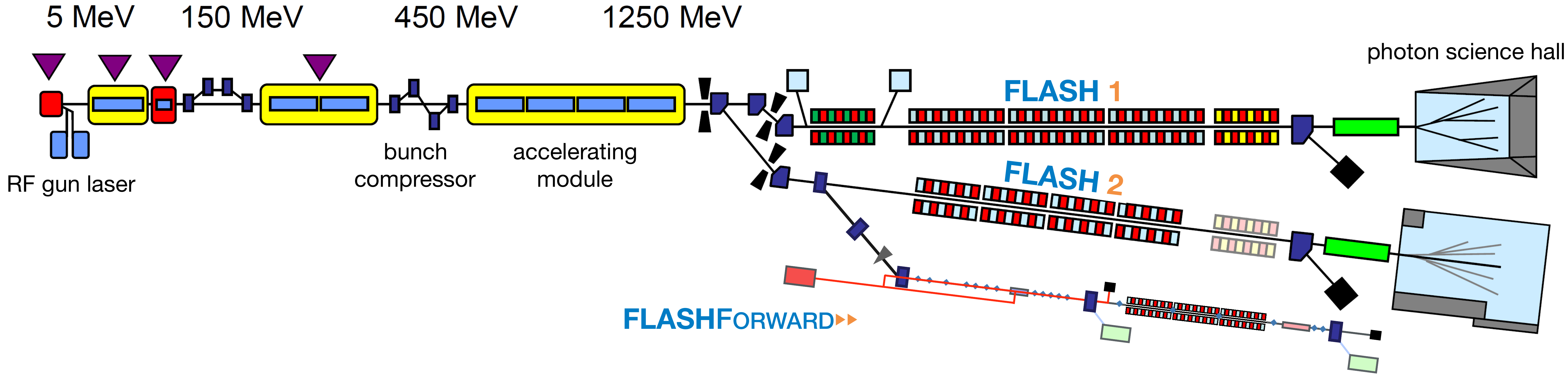}
\caption{\it \footnotesize \label{FIGURE:FLASH} Schematic of the FLASH SRF front-end supplying the FLASHForward experimental beamline with high-average-power electron beams. Also shown are the adjoining FLASH1 and FLASH2 FEL beamlines. The FLASHForward beamline occupies the same experimental interlock area as that of FLASH2.}	
\end{figure}

\begin{itemize}
\item \textbf{Double-bunch generation methods}. It is possible to generate double-bunch structures at FLASHForward for external injection -- essential for staging of PWFA -- in two ways. The first is at the FLASH photocathode gun, whereby a tunable laser-pulse split-and-delay system is used to generate two laser pulses, and thus electron bunches, with picosecond separation. These two bunches, accelerated within the same RF bucket, must have a final post-acceleration separation of less than a plasma wavelength, thus defining the initial delay. The second method utilises a motorised wedge and side-block system, the so-called `scraper', in the FLASHForward compression section, whereby a bunch with a linear correlation in longitudinal phase space may be bisected in both energy and time \cite{Muggli2010}. By varying the bunch length with the accelerating modules and bunch compressors in the linac then translating the wedge and side blocks to different locations, the bunch lengths and separation between the two bunches can be modified.
\item \textbf{Longitudinal bunch shaping}. The presence of a third-harmonic accelerating cavity in the FLASH linac provides immense flexibility over the first, second, and third derivates of the bunch's energy as a function of time. By modifying the properties of this RF module, the longitudinal current profile of the bunch may be manipulated. For example the current profile of the bunch may be modified into a triangular shape, important in order to maximise the acceleration of bunches in PWFA \cite{Piot2012}. This property has a similar importance for the scraper method of double-bunch generation, as the current profile may be manipulated away from a Gaussian profile whilst maintaining a linear energy/time correlation.
\item \textbf{Differential pumping}. The FLASHForward central interaction chamber as well as the compression and final-focussing sections are all equipped with differential pumping stages providing a gradient from high pressure at the plasma chamber to the FLASH vacuum further upstream. This system compensates for the vacuum degradation caused by gas loading of the plasma cell, pumping down from up to 0.1~mbar in the central interaction chamber to the required $10^{-9}$~mbar at the transition to the FLASH vacuum in only 10~m of beamline. Due to this functionality, FLASHForward plasma cells may be windowless, thus eliminating scattering-based emittance degradation upon entering and exiting the plasma cell.
\item \textbf{Megahertz repetition rates}. The FLASH superconducting accelerating modules are capable of operating at MHz frequencies with \SI{}{\bf \micro}s bunch spacing, providing bunch trains to FEL users at up to 30~kW average power. These accelerating modules are capable of sending the same bunch trains to FLASHForward for high-average-power PWFA studies -- a functionality unavailable to any other current or planned PWFA facility. By utilising these bunch trains of variable frequency and power the limits of PWFA's operational ability will be probed, testing the applicability of plasma wakefield schemes to the highest intensity future photon-science and high-energy-physics facilities.
\item \textbf{Femtosecond-level longitudinal diagnostics}. In 2019, the FLASHForward beamline will be extended beyond the current design to incorporate an X-band (11.998~GHz) transverse deflection structure for femtosecond-scale longitudinal phase-space reconstruction \cite{MarchettiIPAC}. This longitudinal diagnostic device will reveal key information about drive beams, allowing energy gain in the plasma to be maximised through bunch shaping. The capability to diagnose witness beams would also yield invaluable insight into the results of acceleration, providing a tool to differentiate between the nuances of distinct injection schemes (thus confirming the validity of particle-in-cell codes), as well as an additional resource in optimising the system for FEL gain. No previous PWFA facility has benefitted from the functionality of an X-band TDS downstream of the plasma cell to diagnose both drive and witness bunches, giving FLASHForward a unique experimental asset.
\item \textbf{Laser-to-electron-beam timing}. In 2020 the timing system of the FLASHForward electron beamline will undergo an upgrade. This will bring the synchronisation of the electron beam with the 25~TW, 25~fs, 10~Hz titanium-sapphire CPA laser system to $\mathcal{O}(10~{\rm fs})$~rms \cite{Schulz2015}. This exquisite level of synchronisation will be realised via a pulsed optical reference signal (rather than electronic RF signals) distributed throughout the FLASH facility, derived from the repetition rate of a femtosecond mode-locked laser oscillator. This level of synchronisation between laser pulses and the electron beam driver will be essential for laser-triggered injection schemes and transverse laser diagnostics.
\end{itemize}

\subsection*{Beamline overview}

\begin{figure}[b]
\centering
\includegraphics[width=\columnwidth]{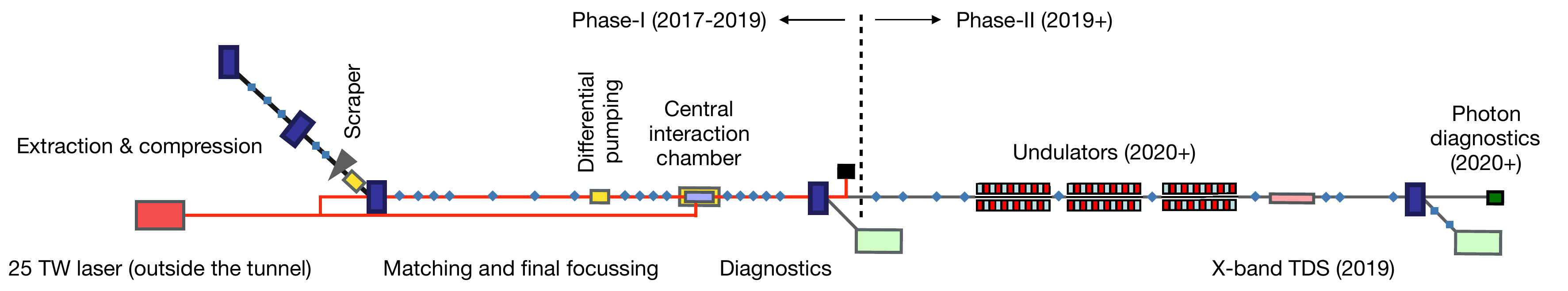}
\caption{\it \footnotesize \label{FIGURE:FLASHForward} Schematic of the FLASHForward experimental beamline, highlighting major sections and components.}	
\end{figure}

The FLASH electron accelerator provides bunches to FLASHForward with sufficient quality to drive an FEL. These bunches have energies up to 1.25~GeV, $\sim$0.1$\%$ uncorrelated energy spread, $\sim$2~\SI{}{\bf \micro}m normalised transverse emittance, variable lengths from 50--5000~fs, bunch charges of 0.1--3.0~nC, and bunch spacings on the sub-\SI{}{\bf \micro}s level. These electron bunches are capable of acting as the driver for the plasma wakefield with peak currents ranging up to $\sim$2.5~kA.

Upon extraction from the FLASH linac and FLASH2 FEL beamline, the electron bunches traverse the FLASHForward beamline. This beamline, shown schematically in Fig.~\ref{FIGURE:FLASHForward}, consists of seven sections: extraction and compression, matching and final focussing, central interaction area, initial diagnostics, undulators, X-band TDS, and the final dump. At present the electron beamline is complete up to the end of the initial diagnostics section but will be extended in 2019, to facilitate the installation of the X-band TDS. The beamline will culminate in the 2020s with the implementation of a series of Tesla Test Facility (TTF) undulator modules.

The extraction and compression section is designed to extract bunches and provide a 4~m transverse separation from the FLASH2 FEL beamline, as well as providing variable bunch compression in addition to that of the bunch compressors in the linac. The trajectory change from the extraction dipole is corrected by two dipoles utilised in the beam optics design to close dispersion to both first and second order. The transfer-matrix element relating the energy and time domains, $R_{56}$, can be varied between \num{-5} and 4~mm. It is in this section that the `scraper' is located, utilising a mask and wedge system to remove certain parts of a linearly chirped beam.

Once parallel with the FLASH2 beamline, the beam interacts with a series of focussing optics intended to focus the electron beam down to 5~\SI{}{\bf \micro}m transverse size at the interaction point. It is expected that the spatial and angular transverse jitter will be minimal at this point, <10~\SI{}{\bf \micro}m and <0.5~mrad respectively. The differential pumping stages, implemented to prevent critical vacuum degradation at the intersection with the FLASH2 vacuum, are housed in this final focussing section as well as the compression section.

The central interaction chamber is the heart of the beamline, containing the plasma capillary i.e. the interaction point of the laser and electron beams. Housed within this chamber and mounted on the same plasma-cell baseplate are a series of transition-radiation and scintillating screens, as well as knife-edge diagnostics for spatial alignment of the laser and electron beams. The central interaction chamber is also pumped in two sections: the top section contains the plasma cell and diagnostics; the bottom contains a hexapod positioning system on which the baseplate is placed, providing precisely controlled rotational and translational movement of the plasma cell and diagnostics.

The multiple side ports incorporated into the design of the central interaction chamber provide additional access for transverse laser pulses, intended to probe laser/electron/plasma interactions, as well as clear lines of sight for optical diagnostic systems. The plasma sources installed within the chamber can operate with a range of gas species, up to plasma electron densities of $\mathcal{O}(10^{18}~{\rm cm^{-3}}$) for windowless cells and $\mathcal{O}(10^{20}~{\rm cm^{-3}}$) for capped capillaries. The geometry of the base plate can accommodate multiple plasma cells of up to 300~mm in combined length.

The diagnostic system immediately downstream of the plasma cell contains a series of high-strength capturing quadrupoles, designed to capture and transport witness beams of up to 2.5~GeV. Once captured and focussed, the beams interact with a dipole magnet used as a broadband energy spectrometer to measure the energy of the driver and witness bunches.

In summer 2019 the FLASHForward beamline will be extended by $\sim$40~m to incorporate the X-band TDS. The beamline extension will include a series of quadrupoles, both upstream and downstream of the TDS. These are necessary to provide the optics flexibility required to optimise the temporal resolution of the TDS and are placed such that suitably sized gaps for the TTF undulator modules exist. A screen station, similar to those in the compression and matching sections, will be used to reconstruct both the slice emittance in one transverse plane (down to nm resolution) and the geometric emittance (by way of a quadrupole scan method) in the other. A second broadband electron spectrometer will be implemented towards the end of the beamline to provide the dispersion necessary to determine the longitudinal phase space of driver and witness bunches facilitated by the X-TDS.


\section{Research goals}\label{sec:research}

The FLASHForward facility, as described in Sec.~\ref{sec:FLASHForward}, is designed to investigate a core feature of PWFA research -- stable and reproducible GV/m-gradient acceleration of high-frequency, high-quality beams for future applications. This  goal is broken down into a series of core studies, pioneered by the flagship internal and external injection experiments; the so-called X-1 and X-2 experiments, respectively. The results obtained from these studies over the coming years will feed directly into the main goal of FLASHForward: the demonstration of FEL gain from beam-driven PWFA. A series of ancillary experiments, running in parallel to and feeding into the core studies, is also of immediate interest to the field. This broad scientific programme is outlined in the following section.

\subsection{X-1: High-brightness beam generation}

\begin{figure}[b]
\centering
\includegraphics[width=0.7\textwidth]{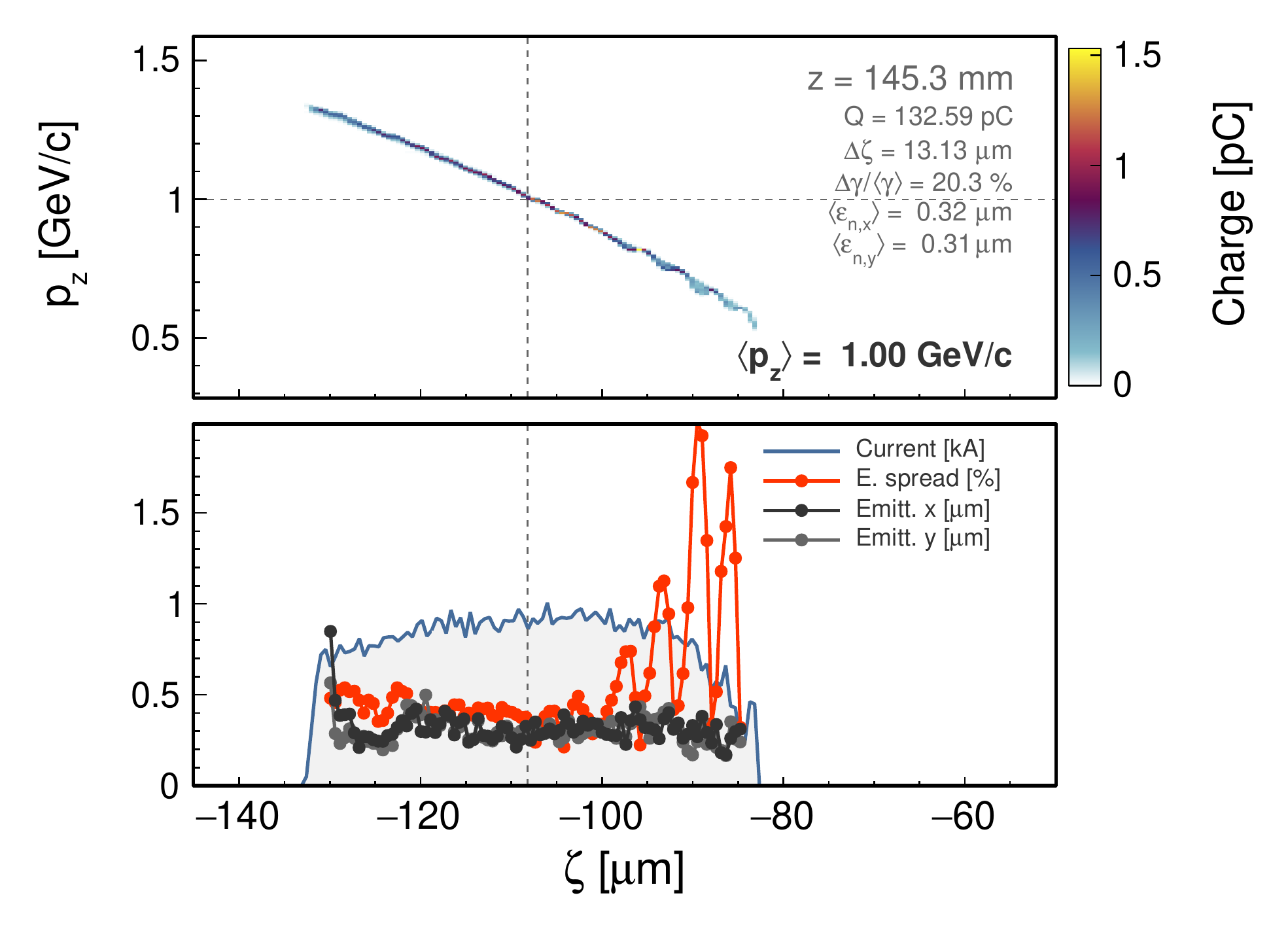}
\caption{\it \footnotesize \label{FIGURE:internal} (Top) Longitudinal phase-space and (bottom) slice energy spread (red), normalised transverse emittance (black, grey), and peak current (blue) of the witness bunch after a propagation distance of $\sim$15~cm simulated in {\tt OSIRIS 3D} \cite{osiris1,osiris2,osiris3}. A typical FLASHForward 1 GeV drive beam with 0.1$\%$ relative energy spread, $\sigma_z=25$~\SI{}{\bf \micro}m~rms length, $\sigma_{x,y}=6$~\SI{}{\bf \micro}m~rms width, 2~\SI{}{\bf \micro}m normalised transverse emittance, a charge of 500~pC, and a peak current of 2.4~kA was chosen to drive the wake. The plasma electron density at the peak of the ramp was $4\times10^{17}$~cm$^{-3}$ (ten times higher than that of the flat-top) with a downramp length of $\sim$100~\SI{}{\bf \micro}m.}
\end{figure}

In recent years a number of novel beam-driven injection techniques have been proposed \cite{hidding2012,li2013,martinez2013,wittig2015}. One promising approach is based on the concept of injecting electrons from the background plasma by means of controlled wave-breaking during a plasma density down-ramp (DDR) transition \cite{martinez2017}. In the case of FLASHForward this particular injection method has the benefit of requiring lower peak currents than other methods, e.g. beam-induced ionisation injection, adding additional flexibility to beam-shaping at the expense of peak current. Results of 3D particle-in-cell (PIC) simulations, such as the example illustrated in Fig.~\ref{FIGURE:internal} for typical FLASHForward experimental conditions, suggest promising results, demonstrating the potential to generate high-current, GeV electron bunches with uncorrelated energy spreads of $\sim$0.5$\%$ and small transverse normalised emittance (0.3~\SI{}{\bf \micro}m). It is beams with these transverse emittance, an order of magnitude lower than the FLASH bunches used to generate them, which are of particular interest to the FEL community as they may boost the 6D brightness of the machine by approximately an order of magnitude (two orders of magnitude up via a reduction in the transverse emittance and one down via an increase in the projected energy spread).

At FLASHForward these downramp transitions will be generated by localised ionisation of a doped gaseous species via a laser propagating orthogonally to the electron-beam direction. This approach is feasible at FLASHForward due to the pre-existing 25 TW laser system and the exquisite laser-to-electron-beam synchronisation of the machine. These density spikes are anticipated to have lengths on the tens-of-\SI{}{\bf \micro}m-scale, up to an order of magnitude shorter than those generated using hydrodynamic methods. The exact properties of this transition strongly influence the injected beam parameters \cite{martinez2017}. The tunable nature of these downramp profiles, thanks to the high level of control over laser parameters, makes DDR an appealing avenue of exploration for photon science applications and forms the first pillar of research at FLASHForward.

\subsection{X-2: High-quality plasma booster}

\begin{figure}[b]
\centering
\includegraphics[width=0.65\textwidth]{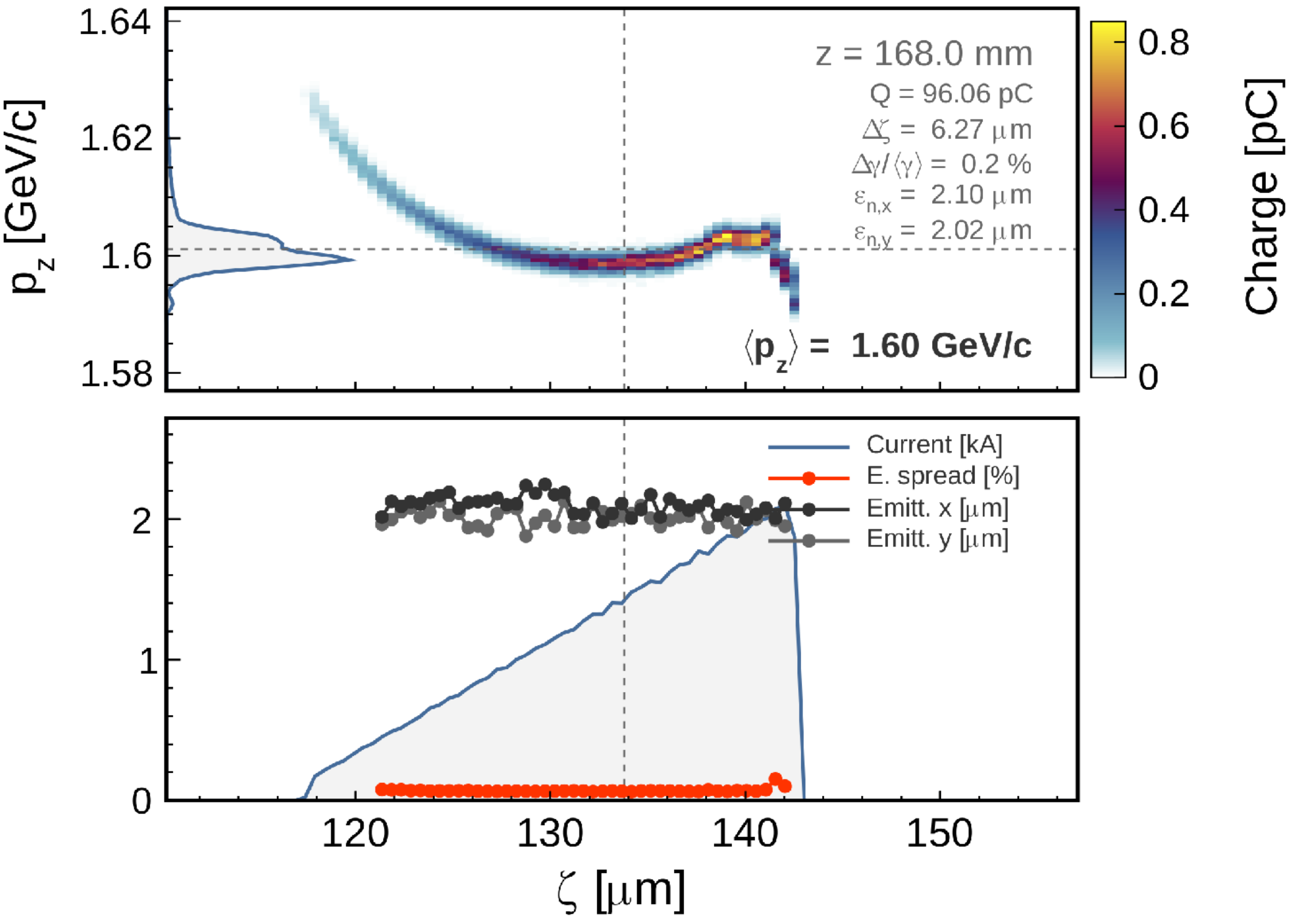}
\caption{\it \footnotesize \label{FIGURE:external} (Top) Longitudinal phase-space and (bottom) slice energy spread (red), normalised transverse emittance (black, grey), and peak current (blue) of the witness bunch after a propagation distance of $\sim$10~cm simulated in {\tt HiPACE} \cite{HiPACE}. A typical FLASHForward 1 GeV drive beam with 0.2$\%$ relative energy spread, $\sigma_z=40$~\SI{}{\bf \micro}m~rms length, $\sigma_{x,y}=5$~\SI{}{\bf \micro}m~rms width, 2~\SI{}{\bf \micro}m normalised transverse emittance, and a charge of $\sim$500~pC was bisected and trimmed by the scraper, with the parameters optimised to provide a loaded wakefield at the point of the witness. The plasma electron density of the flat-top acceleration region was $5\times10^{16}$~cm$^{-3}$.}
\end{figure}

Due to the GV/m accelerating gradients available to PWFA schemes, their application to future high-energy-physics facilities is clearly advantageous. In one such concept the plasma cell would act as an energy booster for a particle bunch generated from, for example, a conventional photocathode gun. This witness bunch would be externally injected immediately behind a drive bunch, separated by less than a plasma wavelength, and would be boosted in energy. The expansion of this concept to multiple stages has been foreseen, in which multiple plasma-cell stages are used to boost electron and positron bunches to the TeV energy scales required for future linear collider facilities \cite{Adli2013}. However, this requires that the favourable bunch qualities of the injected bunch are maintained over distances necessary for substantial energy gain. Studies have been performed at FACET demonstrating conservation of uncorrelated energy spread \cite{FACET2} through beam loading of the wake. However, other favourable beam properties such as emittance -- essential for collider facilities in order to maximise final focus luminosities -- must also be conserved. The FLASHForward X-2 experiment \cite{Libov2018} is designed to answer these questions.

In order to inject bunches into plasma externally, a driver-witness pair is generated using the previously described `scraper' method. Simulated bunches, generated at the FLASH gun in {\tt ASTRA} \cite{ASTRA} were then propagated through the FLASH and FLASHForward lattice in {\tt elegant} \cite{elegant} and are used to produce a double-bunch structure most conducive to quality preservation in the plasma. The result of such an optimisation process can be seen in Fig.~\ref{FIGURE:external}, in which a bunch has been accelerated from 1 to 1.6~GeV in 168~mm of plasma, representing an accelerating gradient of 3.75~GV/m, with negligible emittance growth and energy spread preservation. Experimental realisation of this at FLASHForward would require preservation of beam quality during acceleration, extraction from the plasma module, and transport to interaction or diagnostics regions.


\subsection{Ancillary experiments}
In addition to the two flagship experiments, the scientific programme of FLASHForward includes a diverse range of ancillary experiments running in parallel to X-1 and X-2. These studies are closely linked to the core studies, either occurring as a direct result of progress made within them or contributing crucial steps to their realisation. In all cases, however, they are of individual interest to the wider community and as such the results are directly applicable to other experiments in the field. In many cases these experiments are made possible due to the unique infrastructure available to FLASHForward (outlined in Sec.~\ref{sec:FLASHForward}). Some areas of particular interest are: 

\begin{itemize}
\item \textbf{Plasma cell characterisation}. As previously mentioned, quality-preserving extraction of witness bunches from a plasma module is essential for the future prospects of staging. It is therefore necessary to suitably tailor and measure the plasma-density profiles at the exit of the cell in order to optimise the witness beam release from plasma. At FLASHForward's diagnostics laboratory -- situated directly above the electron beamline tunnel, hosting a laser path with identical path lengths and optics to that of the laser transport to the interaction point in the tunnel -- the plasma profiles in these windowless cells (capable of plasma generation via an ionisation laser or electric discharge) are characterised using a range of techniques including spectroscopy \cite{spectroscopy} and group dispersion velocity \cite{GVD}, with the results then benchmarked against magneto-hydrodynamic codes.
\item \textbf{Active plasma lens development}. PWFA accelerated beams have, in general, favourably small transverse emittance and durations. However, upon exit from a plasma cell these qualities will result in rapid beam blow-up in a drift length. A strong magnetic field is therefore required immediately downstream of the plasma exit in order to capture the beam and prevent emittance increase. Conventional permanent quadrupoles are often used for this purpose; however, they have fixed strengths and require multiple magnets in order to focus in both planes. The FLASHForward collaboration has therefore carried out research into plasma-based focussing devices, recently demonstrated to focus in both planes with kT/m strengths \cite{APL1} without degrading emittance \cite{APL2}.
\item \textbf{Transformer ratio optimisation}. Through use of the third-harmonic RF cavity in FLASH, as detailed in Sec.~\ref{sec:FLASHForward}, the current profile of the beam can be shaped to non-Gaussian distributions, thereby testing limits of the transformer ratio, i.e. the ratio of the peak electric field generated by the driver and the peak accelerating field experienced by the witness, theoretically limited to two for Gaussian bunch profiles in the linear regime. The breaking of this limit has been recently demonstrated at DESY \cite{Loisch2018}. However, it is essential to reproduce then maximise these results at a GeV-class machine to test the efficacy of the principle at future facilities.
\item \textbf{Hosing characterisation and mitigation}. Identified by Whittum {\it et al.} in the early 1990s \cite{hosing1a}, the hose instability remains a long-standing challenge for PWFAs. Much effort has been invested in recent years into developing an accurate analytical model of hosing \cite{hosing1b,hosing1c}, with simulations subsequently suggesting that e.g. an initial beam-energy spread and/or large transverse emittance will mitigate the hosing instability \cite{hosing2,hosing3}, thus stabilising propagation of the drive beam over large distances. In the context of FLASHForward, this hosing instability will be artificially generated by introducing a centroid offset in the drive beam, observed and characterised using the suite of diagnostics available to FLASHForward, then mitigated through careful tuning of other drive-beam properties.
\item \textbf{Plasma-based dechirpers}. A challenge of plasma-based concepts operating with GV/m gradients is the development of the longitudinal phase-space of the beam, accelerated in an environment that may imprint a large linear energy-time dependency on the beam up to the GeV/mm level. This large negative chirp at the exit of the plasma section will halt FEL gain or lead to a beam-size increase limiting luminosity in HEP experiments and must therefore be mitigated with gradients stronger than those currently available to other state-of-the-art techniques e.g. dielectric devices. Efficacy of a plasma-based technique with GV/mm/m dechirping strengths has recently been demonstrated at FLASHForward \cite{dechirping}, marking the first experimental result from the electron beamline.
\item \textbf{High-average-power PWFA}. The deployment of PWFA techniques at future facilities requires operation at MHz frequencies in order to be able to compete with conventional accelerator technology. Due to the superconducting RF cavities of the FLASH linac, bunch trains with \SI{}{\bf \micro}s spacing and average powers $\mathcal{O}(10~{\rm kW})$ may be provided to FLASHForward for high-average-power PWFA research. These average powers, orders of magnitude higher than other state-of-the-art PWFA facilities, will be used to probe the applicability of PWFA techniques to the next generation of photon-science and high-energy-physics machines. Further details on the conceptual design are given in Sec.~\ref{sec:HAP}.
\end{itemize}

\subsection{X-100: PWFA FEL gain}
A major scientific goal of FLASHForward is to examine whether beams from both X-1 and X-2 are capable of driving FEL gain at the few-nanometre scale. In each case, the technical progress and hardware results of both core experiments and ancillary studies will feed into this final goal. This demonstration of FEL gain would be defined as a one-hundred-fold magnification of the self-amplified spontaneous emission (SASE) signal, therefore operating in the exponential regime. A demonstration of this type would be the first major breakthrough en route to a plasma-based tabletop FEL facility and would be of the utmost significance in the field.

A number of the Tesla Test Facility (TTF) undulator modules will be available for implementation at FLASHForward after 2020. A preliminary scan of the parameter space via the Xie formalism for these TTF undulators (K = 1.27, $\lambda_u$ = 27.3 mm) and PWFA beams from particle-in-cell simulations has beam performed \cite{Niknejadi2017}. This shows that PWFA beams are suitable for demonstration of FEL gain, particularly if strong focusing is implemented. In addition to the 3D effects considered via the Xie formalism, strong slippage effects i.e. the deceleration of an electron from one pre-defined slice to another must be avoided. Therefore the PWFA bunch lengths must be longer than the slippage length. For external injection parameters, the minimum bunch length is $\sim$5 fs and for DDR parameters $\sim$9 fs. These initial criteria have been met in the preliminary simulations. Full 3D simulations are planned to investigate this scheme further.


\section{X-3: High-average-power at FLASHForward}\label{sec:HAP}

As outlined in Sec.~\ref{sec:research}, the success of both X-1 and X-2 experiments could drastically reduce the size of future facilities. However, despite the rapid and promising evolution of laser- and particle-driven plasma based accelerators (PBA), neither has as yet been able to provide the high-average-power afforded by superconducting RF (SRF) technology, thus drastically limiting the applicability of both schemes for future facilities. LWFA is currently limited by the available repetition rate of high-intensity, high-power laser drivers, presently preventing them from operating at repetition rates beyond the Hz level. Although schemes in which high-intensity, Joule-level lasers are able to operate at kHz rates have been envisaged, these are years away from experimental realisation. For PWFA machines, the limitation in repetition rate arises from the conventional accelerator that provides the drive beam. However, unlike the case of LWFA schemes, high repetition rate, high-average-power conventional facilities exist \cite{FLASH,XFEL} and are able to provide beams to investigate the efficacy of PWFA as drivers for future collider and FEL installations.

\begin{figure}[h]
\centering
\includegraphics[width=0.5\columnwidth]{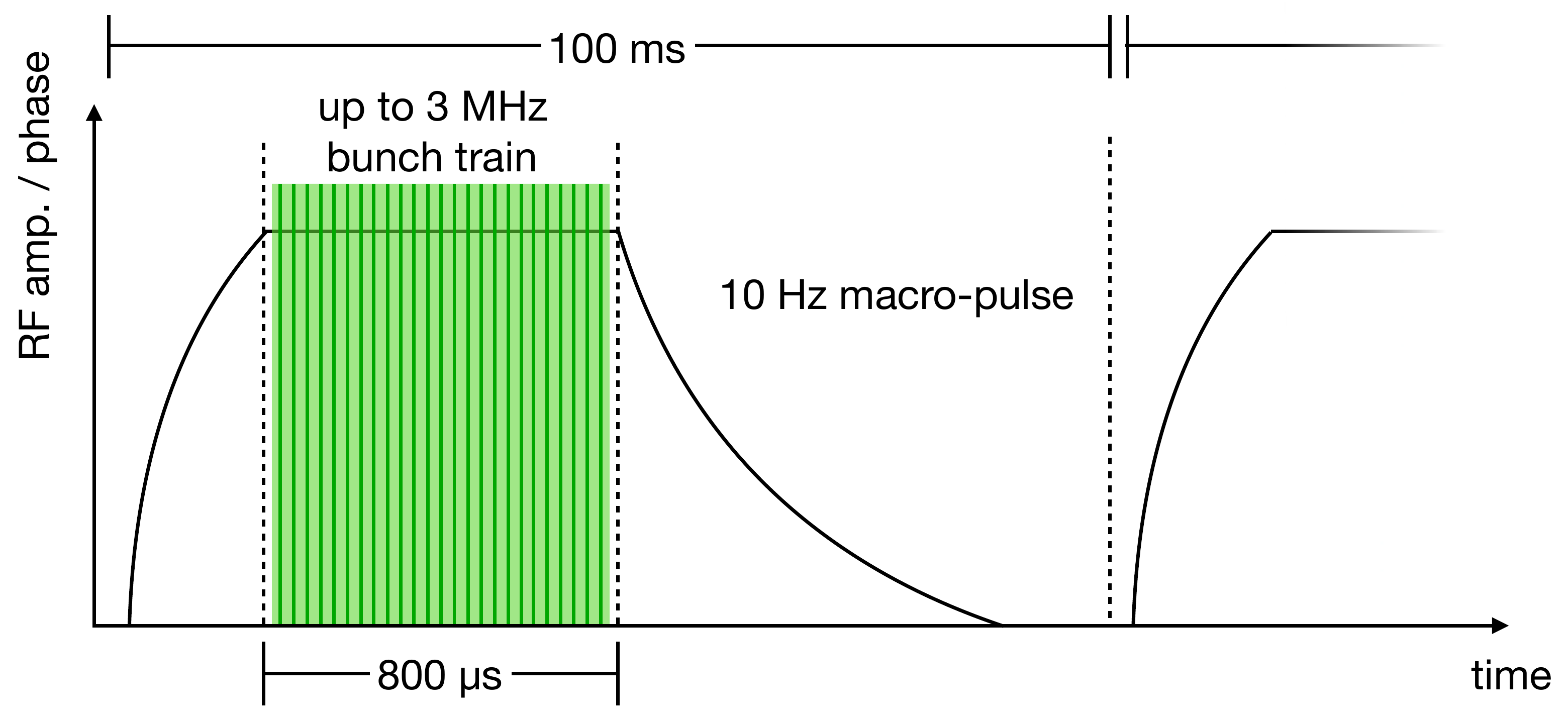}
\caption{\it \footnotesize \label{FIGURE:bunchstructure} Schematic of the 10~Hz macro-pulse and MHz bunch train structure available to FLASHForward for high-power experimentation.}
\end{figure}

As described in Sec.~\ref{sec:FLASHForward}, initial acceleration of electrons at FLASHForward is carried out by a series of SRF stages using niobium accelerating cavities. The FLASH gun and linac modules operate with a 10~Hz macro-pulse structure, providing 1.25~GeV, 1~nC electron bunches at up to 3~MHz micro-pulse frequencies. This bunch structure can be seen in Fig. 5, illustrating the maximum 800~\SI{}{\bf \micro}s long SRF flat-top pulse. At full capacity, the SRF modules can accommodate 2,400 bunches per macro-pulse at 3~MHz micro-pulse frequencies, thus providing up to 24,000 bunches per second to the FLASHForward experimental area.

\begin{figure}[b]
\centering
\includegraphics[width=0.6\columnwidth]{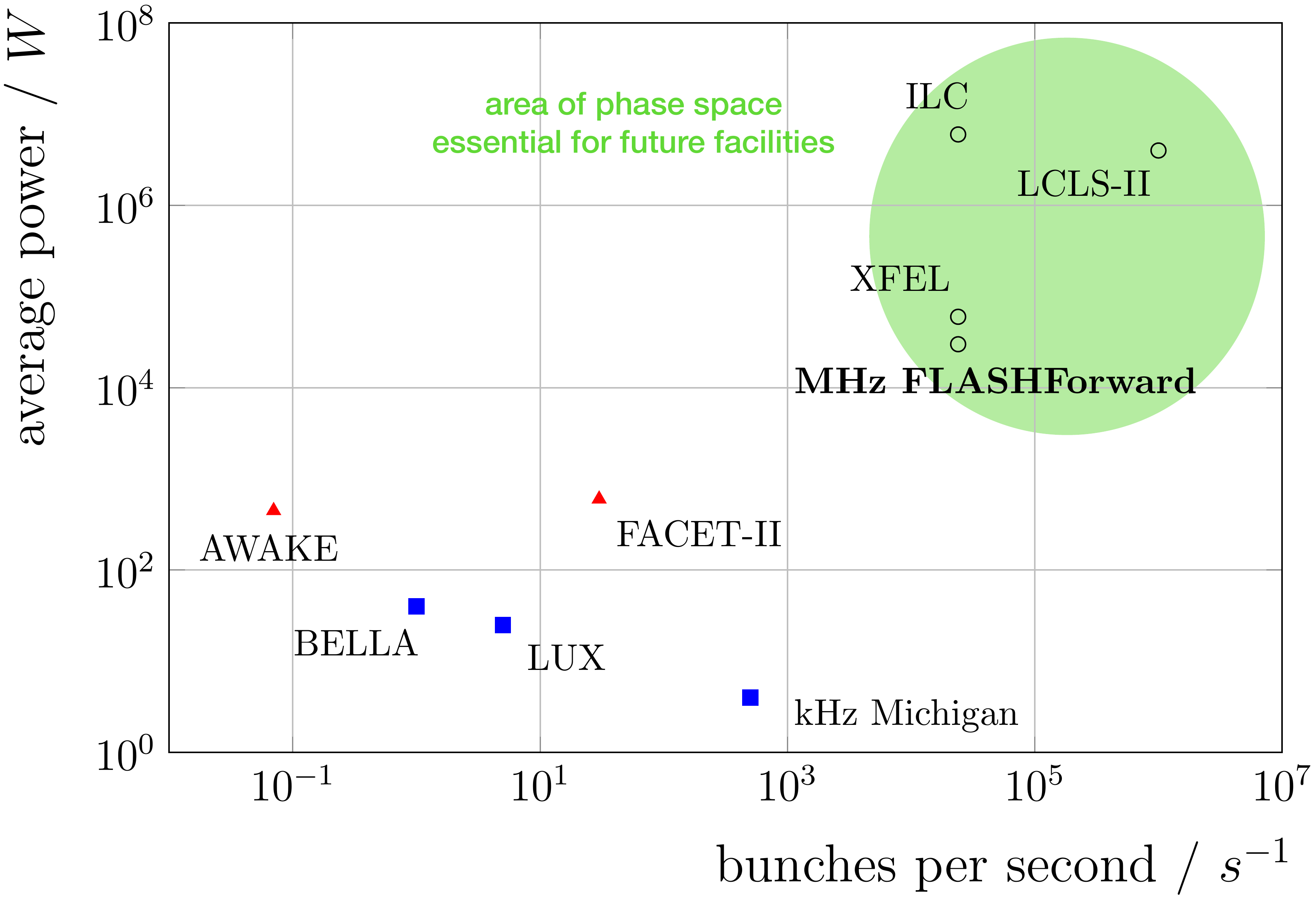}
\caption{\it \footnotesize \label{FIGURE:powerplot} Average-power capabilities (as a function of the number of bunches per second their drivers can deliver) of state-of-the-art LWFA (blue squares) and PWFA (red triangles) facilities. The white circles represent the state-of-the-art or next generation of photon science and collider facilities. By implementing certain infrastructural changes to FLASHForward, PWFA research would move into the region of phase space necessary for future facilities.}
\end{figure}

Figure~\ref{FIGURE:powerplot} shows a handful of cutting edge LWFA and PWFA facilities, the average power of which are displayed as a function of the number of bunches per second their front ends are capable of supplying. In order to pursue the application of PBA technology to future collider and photon-science facilities, its functionality must be demonstrated at the high-average-powers currently available via conventional accelerator technology. At full capacity, the FLASH bunch-train structure corresponds to 30~kW of average power, orders of magnitude higher than drivers available at other state-of-the-art PBA facilities. This high-power functionality makes FLASHForward the only plasma-wakefield facility in the world with the capability to develop, explore, and benchmark high-average-power plasma-wakefield research.

However, in order to fully utilise this high-power capability the current FLASHForward infrastructure must be expanded. Implementation of the concepts outlined in this section will facilitate this expansion. These are centred around i) proof-of-principle sub-\SI{}{\bf \micro}s probing of the PWFA recovery time, ii) generation of PWFA at MHz frequencies, iii) separation of high-power driver and witness pairs, iv) state-of-the-art capillary lifetime and cooling studies, and v) the development of new diagnostic techniques for the measurement of MHz PWFA. 

\subsection{Plasma recovery time}\label{SEC:recovery}
Almost all high-energy PBA machines around the world are restricted to operation in the range of 1--100~Hz. However, some low-energy machines have demonstrated LWFA at up to the 1~kHz level \cite{kHzLWFA}. The replenishment of plasma via a second ionisation of fresh gas, such that the two wakefields result in an identical accelerating effect -- the plasma recovery time -- is therefore given an upper bound of 1~ms as defined by the 1~kHz repetition rate of these low-energy, high-frequency machines. In order to test the applicability of PWFA schemes to future facilities, probing of the plasma recovery time down to the sub-\SI{}{\bf \micro}s level -- three orders of magnitude lower than the current limit -- is essential.

\begin{figure}[h]
\centering
\includegraphics[width=0.8\columnwidth]{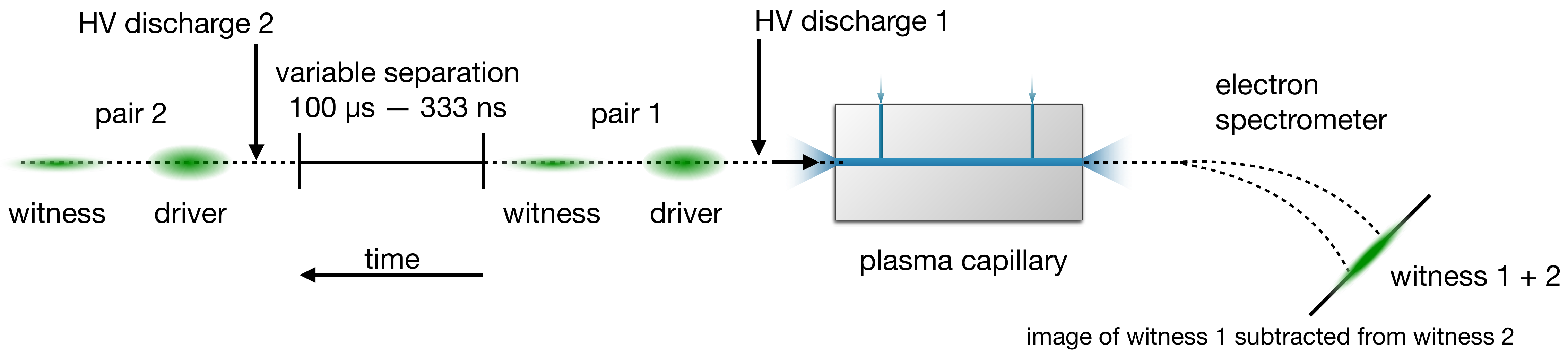}
\caption{\it \footnotesize \label{FIGURE:twopairs} Schematic of the proof-of-principle effective MHz experimental setup illustrating two driver-witness bunch-pairs with variable separation interacting with two distinct plasmas generated by consecutive high-voltage discharges.}
\end{figure}

Such an investigation at FLASHForward requires minimal modifications to the infrastructure. The concept is based on the consecutive creation of plasma via a high-voltage discharge with separation greater than a microsecond. Immediately after each plasma creation an electron driver-and-witness bunch-pair (as illustrated in Fig.~\ref{FIGURE:twopairs}) would be injected into the plasma. The acceleration effect experienced by the first witness bunch would then be compared to that of the second for a range of offsets between the time of the first discharge and the second.

In order to create two back-to-back plasma generations with sub-\SI{}{\bf \micro}s separations, the FLASHForward experimental infrastructure must be expanded to include a second high-voltage discharge unit. This will be identical to the first and is necessary in order to measure the time it takes for the plasma to return to a homogeneous state in time for the subsequent second electron bunch-pair. By operating with two identical and independently controlled thyratrons, two ionising discharges may be fired through the gas at the same sub-\SI{}{\bf \micro}s separation producible by the FLASH SRF driver. By matching the separation in time of these two discharges to that of the two incident electron bunch pairs the plasma recovery time will be investigated in order to establish the minimum bunch separation required to give comparable PWFA results at an effective MHz repetition rate. This first concept will provide a necessary building block for subsequent studies, in particular and most importantly, expanding this process to higher, and therefore more applicable, average powers.

\subsection{High-frequency plasma generation}
In order to scale up the proof-of-principle two-thyratron experiment to multi-bunch trains, resulting in MHz-rate PWFA of hundreds of bunches, a new type of high-voltage discharge unit must be developed. These bunch trains of increasing average-power, again with intra-bunch separation on the sub-\SI{}{\bf \micro}s-level, will be used to probe the reproducibility of PWFA processes at scales required for implementation at future facilities, answering questions such as when a steady state of reproducible acceleration is reached. Successful operation with hundreds of bunches would confirm that high-average-power PWFA is feasible.

In order to meet this goal, it is clear that a more flexible and efficient discharge switch system is required beyond, say, simply stacking several thyratron units. Much effort has recently been invested in the world-leading Eu-XFEL facility at DESY \cite{XFEL}, with FEL user operation having begun in 2017. One of the hardware developments necessary for deflection of designated bunches in a MHz bunch train into separate beamlines is that of a fast kicker magnet operating at the sub-\SI{}{\bf \micro}s level. Such magnets require rise and fall times on the ns-level with flat tops of tens-of-ns in length. These fast kickers are switched by a type of metal-oxide-semiconductor field-effect transistor (MOSFET) operating at MHz rates and voltage amplitudes on the few-kV scale. The summing of these transistors to provide tens-of-kV pulses would allow for complete ionisation of the outer electrons of most gas species used in PWFA experiments, delivering the perfect method for high-frequency (re)ignition of plasma with variable frequency and separation.

Once the hardware development of the MOSFET and its subsequent incorporation into the FLASHForward infrastructure are complete, the results of the double-thyratron proof-of-principle experiment may be benchmarked against those of the MOSFET. The number of consecutive bunches in a single macro-pulse bunch-train will then be increased from two to the current 60~W average-power limit of the FLASHForward facility (currently defined by radiation safety limits but which can in future be upgraded). This corresponds to, for example, 200 electron bunches of 1~GeV energy and 0.3~nC charge at a macro-pulse-repetition rate of 1~Hz.

\subsection{Capillary lifetime and cooling}

The transfer of energy from the driver to the wake then to the witness bunch is not 100$\%$ efficient, with most of the unused energy transmitted to the plasma in the form of heat. As an example, assuming the process is 50$\%$ efficient would lead to approximately 1~kW/cm of power dissipation for a 15~cm capillary at the full FLASHForward average power. Magneto-hydrodynamic codes such as {\tt FLASH} \cite{FLASHcode} must be used to simulate the heat deposition and transfer in a variety of gas species. The subsequent transfer of that heat through the plasma and into the surrounding capillary walls must also be quantified in order to develop a suitable cooling system.

Investigation into the lifetime of plasma capillaries over millions of kHz-rate, high-voltage discharges has shown promising solutions that could be implemented at FLASHForward \cite{Gonsalves}. The heating of capillaries from millions of plasma ignitions and wall erosion from kHz repetition discharges have been studied. However, the studies would need to be expanded to situations with MHz frequencies and plasma heating to the system from drive-beam-power deposition. An experimental test-stand should therefore be developed to benchmark the magneto-hydrodynamic codes simulations experimentally, prototyping various capillary materials such as ceramics, sapphire, diamond, etc. These designs will then be implemented in the FLASHForward experimental setup in order to assess their durability. Once the capillary geometry and material have been optimised, a suitable cooling system will be developed in close collaboration with the expertise and personnel of the DESY Machine Cooling Group (MKK).

\subsection{Separation of driver and witness bunches}
Separation of accelerated witness bunches from depleted drive bunches is a topic of pressing concern to the community at all operational frequencies. The very small beam sizes typical for PWFA will grow exponentially upon exiting the plasma-cell. It is therefore essential to both focus the witness bunch for transport/further study and simultaneously deflect the drive bunch for disposal in as short a longitudinal distance as possible.

\begin{figure}[h]
\centering
\includegraphics[width=0.8\columnwidth]{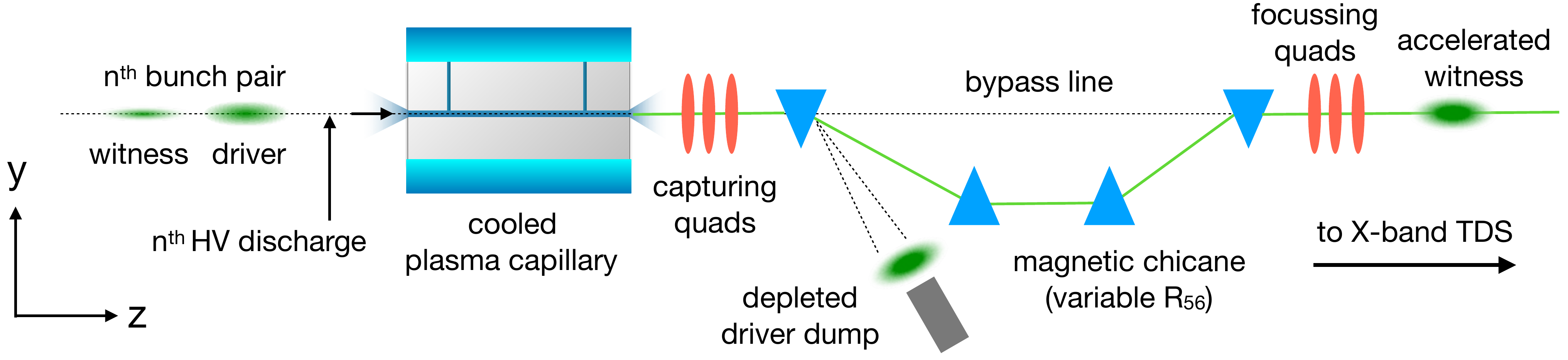}
\caption{\it \footnotesize \label{FIGURE:driverdump} Schematic of the proposed beamline scheme for separation of driver and witness bunches in high-power multi-bunch trains.}
\end{figure}

One possible scheme for implementation at FLASHForward can be seen in Fig.~\ref{FIGURE:driverdump}, whereby a series of high-gradient quadrupoles immediately focuses the accelerated witness and depleted drive bunches into a magnetic chicane. After the drive beam is deflected downwards by the first dipole, a conventional high-power dump will be used to dispose of the beam. The remaining bending dipoles in the chicane would then be used to deflect the witness bunch back onto the original beam path and, depending on the magnitude and linearity of the energy--time correlation of the witness bunches, provide compression to shorter lengths and thus higher peak currents. This modified beamline should include a bypass line providing an unperturbed beam path for conventional low-power, single-bunch operation of FLASHForward.

\subsection{Measurement of MHz acceleration effect}

In order to demonstrate the successful acceleration of witness bunches in multi-bunch trains, a novel diagnostic technique beyond that of the two-bunch subtraction method (i.e. utilising high statistics to subtract the profile of one bunch pair from another within the same ms-level response time of a LANEX-type scintillator screen) must be realised. This technique must be able to identify and characterise the energy profile of each accelerated witness bunch at any point along a multi-bunch train. In the middle of 2019, the FLASHForward beamline will undergo an extension to add a novel dual-polarisation X-band (11.998~GHz) transverse-deflection structure (X-TDS). This state-of-the-art cavity, developed by a collaboration of scientists at DESY, PSI, and CERN \cite{MarchettiIPAC}, will be prototyped at FLASHForward and subsequently utilised to provide a world-first measurement of the longitudinal phase space of an accelerated witness bunch with femtosecond resolution \cite{DArcyIPAC}. Due to the \SI{}{\bf \micro}s-level modulator pulse length driving the cavity and the comparable MHz intra-bunch spacing, the arrival time of the modulator pulse may be varied with respect to the start of the first bunch in a multi-bunch train. This variability would lead to selection and diagnosis of discrete bunches along the train. This flexibility will be utilised to assess PWFA in unprecedented detail across hundreds of bunches within a single macro-pulse.

\begin{figure}[h]
\centering
\includegraphics[width=0.8\columnwidth]{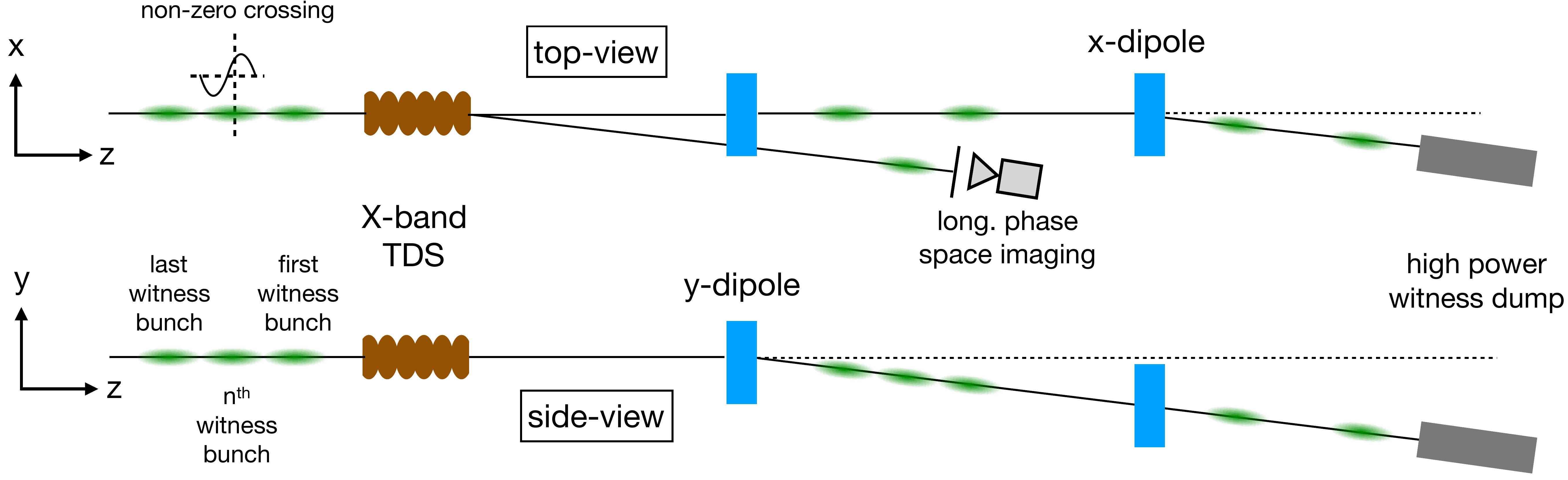}
\caption{\it \footnotesize \label{FIGURE:xband} Schematic of the proposed beamline modifications necessary to utilise the FLASHForward X-TDS for characterisation of MHz PWFA bunch trains.}
\end{figure}

For use with multi-bunch trains, the beamline surrounding the X-band TDS must be modified in order to isolate the desired witness bunch whilst separately dumping the remaining high-power bunch train in a safe way. Due to the special electric fields of the TDS, a bunch will be transversely kicked if it interacts with the RF sine-wave at a non-zero phase. In this case it would provide the necessary force to separate the selected witness bunch from the rest of the multi-bunch train transversely (see Fig.~\ref{FIGURE:xband}). Through the introduction of an additional static dipole magnet, the remaining bunches would then be deflected in $y$, away from the selected bunch and into the pre-existing 150~kW dump originally installed for a now redundant third FLASH FEL beamline. The outlined development requires the beamline optics to be modified but would afford continued parallel operation with FLASHForward when running in low-power, single-bunch mode.

\section{Conclusions}

The FLASHForward facility utilises FEL-quality electron beams from FLASH for plasma-based acceleration research. These beams are ideally suited to PWFA research due to their unique characteristics, such as \SI{}{\bf \micro}m-scale transverse emittance, flexible longitudinal current profiles, and exquisite electron-beam timing. The FLASHForward beamline has been designed to capitalise on these unique properties, with differential pumping sections installed to facilitate windowless capillaries for emittance preservation, a mask in the FLASHForward extraction section to generate double-bunch structures, and a 25~TW, 25~fs laser system with fs-level synchronisation with the electron beam.

The FLASHForward research programme is spearheaded by the flagship internal and external injection experiments, pursuing the applicability of PWFA schemes to future facilities for photon-science and high-energy-physics. In addition to these core studies, a number of satellite experiments are already underway. These ancillary studies exist symbiotically with the core experiments, both aiding them and additionally broadening the FLASHForward programme. The results and progress of all studies at FLASHForward will be utilised to probe for FEL gain from beam-driven PWFA bunches.

However, for the research goals of FLASHForward to be truly applicable to future facilities, high-average-power PWFA operation, at the same order of magnitude as that demanded by photon-science users and high-energy physicists, must be demonstrated. Thanks to the superconducting RF cavities of FLASH, FLASHForward is the only facility in the world capable of pursuing this research at the kW-level, provided a number of essential preconditions can be achieved. Due to the flexible nature of the beamline, as well as future upgrades to the facility such as the X-band TDS, FLASHForward has the potential to make extraordinarily important contributions to the development of plasma-wave acceleration.


\end{document}